\documentclass[11pt]{article}
\usepackage{epsfig}
\usepackage{amssymb}
\usepackage{amsmath}

\newcommand{\id}{{\hbox{{\rm 1}\kern-.24em\hbox{\rm l}}}}

\begin{document}

\noindent
Stockholm\\
USITP 04-7\\
October 2004\\[16mm]

\begin{center}

  {\LARGE \bf{Mutually Unbiased Bases}} \\[3mm]
  {\LARGE \bf{and}} \\[3mm]
  {\LARGE \bf{The Complementarity Polytope}} \\[12mm]

{\large Ingemar Bengtsson\footnote{ingemar@physto.se} 
\ and \
\AA sa Ericsson\footnote{asae@physto.se}} \\[2mm]

{\it Fysikum, Stockholms Universitet} \\
{\it S-106 91 Stockholm, Sweden} \\[16mm]

{\bf Abstract} \\[4mm]

\end{center}

\noindent A complete set of $N+1$ mutually unbiased bases (MUBs) forms
a convex polytope in the $N^2 - 1$ dimensional space of $N\times N$
Hermitian matrices of unit trace. As a geometrical object such a
polytope exists for all values of $N$, while it is unknown whether it
can be made to lie within the body of density matrices unless $N =
p^k$, where $p$ is prime. We investigate the polytope in order to see
if some values of $N$ are geometrically singled out. One such feature
is found: It is possible to select $N^2$ facets in such a way that
their centers form a regular simplex if and only if there exists an
affine plane of order $N$. Affine planes of order $N$ are known to
exist if $N = p^k$; perhaps they do not exist otherwise. However, the
link to the existence of MUBs---if any---remains to be found.

\newpage

\begin{center}
  {\large \bf I. \ Introduction}
\end{center}

\noindent Lately there has been an increase of interest in \emph{mutually
unbiased bases---MUBs} for short---in relation to quantum foundations
and quantum information. Two orthonormal bases $\{ |{\bf e}_i\rangle
\}$ and $\{ |{\bf f}_i\rangle \}$ in Hilbert space~$\mathcal{H}^N$ are
said to be mutually unbiased if, for all $i,j$,
\begin{equation} \label{mub}
  |\langle {\bf e}_i|{\bf f}_j\rangle|^2 = \frac{1}{N} \ .
\end{equation}
Complete knowledge of the observable corresponding to one of the bases
implies total ignorance about the outcome of a measurement in the
other basis. This is a finite dimensional analogy to the complementarity 
of position and momentum in the continuous case.

One of the main features about MUBs was proved in the late eighties by
Wootters and Fields~\cite{Wootters2}. It has to do with optimal state
determination of mixed quantum states. If one wants to determine the
$N^2-1$ parameters of an $N\times N$ density matrix, using von Neumann
measurements, one needs $N+1$ different observables. The statistical
errors in the determination are minimized if one chooses observables
corresponding to bases that are all mutually unbiased. It was seen
(and will be seen later in this paper) that this number of MUBs is the
most there can exist; thus $N+1$ MUBs are called a complete set of
MUBs. But, in spite of efforts that have been made, it is not known if
it is possible in general to have a complete set of MUBs.

Ivanovi\'c~\cite{Ivanovic} has found complete sets of MUBs for prime
dimensions, $N=p$. His construction was generalized by Wootters and
Fields~\cite{Wootters2} to prime power dimensions, $N=p^k$. There the
existence of finite fields with $p^k$ elements is crucial, and further
generalizations are difficult to find since no other finite fields
exists. One conjecture is that it is not possible to have a complete
set of MUBs unless the dimension is a prime power. Using the
construction for prime power dimensions one can find $\textrm{min}\{
p_1^{k_1},\dots,p_r^{k_r} \}+1$ MUBs in the dimension
$N=p_1^{k_1}\cdots p_r^{k_r}$, where the $p_i$'s are all distinct
prime numbers~\cite{Zauner} (see also~\cite{Klappenecker,Archer}). This
gives a lower bound on the maximal number of MUBs. But recently it was
proved that one at least sometimes can have more than
this~\cite{Wocjan}; this is for certain square dimensions.

In this paper we observe that a complete set of MUBs forms a convex
polytope in the set of Hermitian $N \times N$ matrices of unit trace.
In fact, as a convex body such a polytope exists regardless of whether
the MUBs exist or not.  The details are explained in section II, and
result in a reformulation of the question of the existence of a
complete set of MUBs. In section III we turn our attention to the
relation between MUBs and finite affine planes. This connection is
central to Wootters' suggestion as to how the MUBs can be used,
namely, to perform state tomography using a Wigner function defined on
finite affine planes of order~$N$~\cite{Wootters1, Wootters4} (see
also~\cite{Wootters3} for an accessible introduction of the main
ideas). We find an elegant condition, using the language introduced in
section II, that tells us exactly how the affine planes enter the
game. More precisely, we show that an affine plane of order $N$ exists
if and only if a regular simplex can be inscribed in our polytope in a
certain way. On the face of it, this has nothing to do with the
existence of MUBs. In section IV we return to the question of when a
complete set of MUBs can be associated to the polytope; in particular
we investigate whether either the polytope or the set of density
matrices is ``skew'' in some sense. We conclude in section V with a
summary.

\begin{center}
 {\large \textbf{II. \ The complementarity polytope}} 
\end{center}

\noindent Instead of looking at state vectors we will study how the
MUB states sit in the set of
density matrices. This is the subset of positive matrices, in the set
of Hermitian matrices with unit trace. If distances squared between
two matrices are defined by
\begin{equation} \label{distance}
  D^2(A,B) = \frac{1}{2} \textrm{Tr}(A-B)^2 \ ,
\end{equation}
then the set of Hermitian matrices with unit trace form an $N^2-1$
dimensional Euclidean space. Choosing the matrix
\begin{equation}
  \rho_\ast = \frac{1}{N} \id
\end{equation}
as the origin, this is a vector space, with scalar product
\begin{equation}
  (A,B) = \frac{1}{4} \left[ D^2(A+B,\rho_\ast) - D^2(A-B,\rho_\ast) \right]
        = \frac{1}{2}  \left[ \textrm{Tr}AB - \frac{1}{N} \right] \ .
\end{equation}
Thus two matrices are orthogonal if~$\textrm{Tr}AB=1/N$.

The pure states $|\Psi\rangle$ are now the projectors
$P=|\Psi\rangle\langle\Psi|$, sitting on a sphere centered at the
origin and with radius $\sqrt{(N-1)/2N}$. This is the outsphere of the
set of density matrices, which in itself is the convex hull of the
projectors.

The projectors corresponding to two states from an ON-basis in Hilbert
space sit at unit distance from each other. Therefore a full basis
$\{P_k^{(l)}\},\ k=1,\dots,N$, forms a regular simplex in an $N-1$
dimensional subspace. This statement follows from
equation~(\ref{distance}) and the equation
\begin{equation} \label{P1}
  \textrm{Tr} P_k^{(l)} P_{k'}^{(l)} = 0 \ ,\quad k\neq k' \ .
\end{equation}
This kind of simplex will sometimes, for clarification, be called a
\mbox{P-simplex}, since we will later encounter other simplicies. Now
assume that we have several bases $\{P_k^{(l)}\}$, where $l$ labels
the basis and $k$ the state in that basis, which are mutually unbiased
with each other. Then (from eq.~(\ref{mub}))
\begin{equation} \label{P2}
  \textrm{Tr} P_k^{(l)} P_{k'}^{(l')} = \frac{1}{N} \ ,\quad l\neq l' \ . 
\end{equation}
This means that projectors belonging to different MUBs are
orthogonal. Thus the simplices from different MUBs sit in orthogonal
subspaces. Every subspace has dimension $N-1$ and the full dimension
of the space of density matrices is $N^2-1$, so the maximum number of MUBs 
is $N+1$~\cite{Wootters2}. Such a set is called a complete set of MUBs.

If we have a complete set of MUBs, their $N(N+1)$ states defines a
convex polytope---\emph{the complementarity polytope}--- in our
$N^2-1$ dimensional space. But it is possible to have a
complementarity polytope for any~$N$. Just take $N+1$ regular
simplices, each with $N$ corners at unit distance from each other. Let
the simplices sit in orthogonal subspaces, and centered at the
origin. The complementarity polytope is the convex hull of the
$N(N+1)$ corners of these \mbox{P-simplices}. However, it will not be
true in general that the corners are matrices with non-negative
spectrum, that is that they are density matrices. The question of the
existence of a complete set of MUBs can now be reformulated: \emph{Is
it possible to arrange the complementarity polytope so that it is a
subset of the set of density matrices?} We need to somehow rotate the
polytope in such a way that all its corners coincide with pure
states. By construction the corners automatically lie on the outsphere
of the set of density matrices, but generally not at pure states,
since the pure states only constitute a small subset of
the sphere (a~$2N-2$ dimensional subset of the $N^2-2$ dimensional
sphere; the numbers match only for $N=2$). Also when the corners 
of the polytope do not correspond to
density matrices, we will denote them $P_k^{(l)}$---they still obey
equations (\ref{P1}) and (\ref{P2})---or sometimes $P_\omega$, with a
collective index $\omega$ that runs between $1$ and~$N(N+1)$.

So the complementarity polytope exists for all~$N$. It is a distinct 
possibility that
a complete set of MUBs does not exist when~$N\neq p^k$. If this is
true something should happen for these $N$, either with the polytope
or with the set of density matrices. Something that makes it
impossible to rotate the polytope so that it fits in the set of
density matrices.  Is it possible to understand why? Now we will go on
and study the complementarity polytope, to see if we can get any
clues.

We will work in the space of matrices with~$\textrm{Tr}M=1$. When
studying the polytope we need distances, and that means traces like
$\textrm{Tr}M^2$ and $\textrm{Tr}M_1M_2$. In quantum mechanics, on the
other hand, the matrices should be positive, $M\geq0$, to describe
states. To assure this the full spectrum is needed, and therefore
knowledge about $\textrm{Tr}M$ and $\textrm{Tr}M^2$ is not
enough. Also the traces $\textrm{Tr}M^3, \dots, \textrm{Tr}M^N$ are
needed to derive the eigenvalues. The requirements coincide for two by
two matrices, and then it is trivial to fit the polytope in the set of
density matrices (i.e. the Bloch ball).

\begin{figure}
        \centerline{ \hbox{
                \epsfig{figure=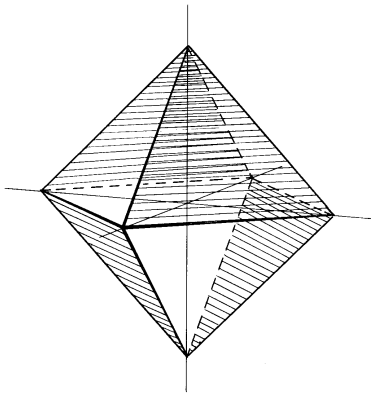,width=5cm}}} 
       \centerline{\parbox{10.4cm}{\caption{The complementarity
        polytope for $N=2$. Four faces have been painted black, 
for reasons that will become clear in section III.} \label{MUB1}}}
\end{figure}

For $N=2$ the complementarity polytope is an octahedron
(figure~\ref{MUB1}). Schematically we can think of it as three one
dimensional orthogonal \mbox{P-simp}\-lices, like
\makebox{
$|$\hspace{-2.6mm}
\raisebox{0.5ex}{$^\bullet$}\hspace{-1.7mm}\raisebox{-1.5ex}{$^\bullet$}
$\, |$\hspace{-2.6mm}
\raisebox{0.5ex}{$^\bullet$}\hspace{-1.7mm}\raisebox{-1.5ex}{$^\bullet$}
$\, |$\hspace{-2.6mm}
\raisebox{0.5ex}{$^\bullet$}\hspace{-1.7mm}\raisebox{-1.5ex}{$^\bullet\,$}.}
When $N=3$ we have four totally orthogonal triangles,
\makebox{
$\triangle$ \hspace{-6.4mm}
\raisebox{-1.2ex}{$^\bullet\;^\bullet$}\hspace{-3.1mm}\raisebox{0.3ex}{$^\bullet$}
$\ \triangle$ \hspace{-6.4mm}
\raisebox{-1.2ex}{$^\bullet\;^\bullet$}\hspace{-3.1mm}\raisebox{0.3ex}{$^\bullet$}
$\ \triangle$ \hspace{-6.4mm}
\raisebox{-1.2ex}{$^\bullet\;^\bullet$}\hspace{-3.1mm}\raisebox{0.3ex}{$^\bullet$}
$\ \triangle$ \hspace{-6.4mm}
\raisebox{-1.2ex}{$^\bullet\;^\bullet$}\hspace{-3.1mm}\raisebox{0.3ex}{$^\bullet$} ,}
when $N=4$ we have five tetrahedra,
\makebox{
$\boxtimes$\hspace{-4.9mm}
\raisebox{-1.2ex}{$^\bullet\;^\bullet$}\hspace{-4.4mm}\raisebox{0.3ex}{$^\bullet\;^\bullet$}
$\boxtimes$\hspace{-4.9mm}
\raisebox{-1.2ex}{$^\bullet\;^\bullet$}\hspace{-4.4mm}\raisebox{0.3ex}{$^\bullet\;^\bullet$}
$\boxtimes$\hspace{-4.9mm}
\raisebox{-1.2ex}{$^\bullet\;^\bullet$}\hspace{-4.4mm}\raisebox{0.3ex}{$^\bullet\;^\bullet$}
$\boxtimes$\hspace{-4.9mm}
\raisebox{-1.2ex}{$^\bullet\;^\bullet$}\hspace{-4.4mm}\raisebox{0.3ex}{$^\bullet\;^\bullet$}
$\boxtimes$\hspace{-4.9mm}
\raisebox{-1.2ex}{$^\bullet\;^\bullet$}\hspace{-4.4mm}\raisebox{0.3ex}{$^\bullet\;^\bullet$},}
and so on for any~$N$. All the edges in the polytope are extremal,
except when $N=2$. A face of the polytope is obtained as the convex
hull of a set of corners, where---to ensure that the face really
belongs to the boundary of the polytope---at least one corner from
each \mbox{P-simplex} is not included. We will be interested in the
faces formed by taking exactly one corner $P_\omega$ from each
\mbox{P-simplex} (in the case $N=3$ this is for example
\makebox{
$\triangle$ \hspace{-3.7mm}\raisebox{0.3ex}{$^\bullet$}
$\triangle$ \hspace{-3.7mm}\raisebox{0.3ex}{$^\bullet$}
$\triangle$ \hspace{-3.7mm}\raisebox{0.3ex}{$^\bullet$}
$\triangle$ \hspace{-3.7mm}\raisebox{0.3ex}{$^\bullet$} )}
There are $N^{N+1}$ such faces and these faces are themselves $N$
dimensional regular simplices. We call them point faces (in the next
section, we will see that they are related to Wootters' phase
points~\cite{Wootters4}). To each point face $\alpha$ we will
associate a ``point face operator''
\begin{equation}
  A_\alpha = \rho_\ast + \sum_{P_\omega \in \alpha} (P_\omega - \rho_\ast)
           = \sum_{P_\omega \in \alpha} P_\omega - \id \ . \label{7} 
\end{equation}
For these operators we have
\begin{equation}
  \textrm{Tr}A_\alpha = 1 \ \ \textrm{ and } \ \ \textrm{Tr}A_\alpha^2 = N \ . 
\end{equation}
The second equality follows from
\begin{equation}
  \big( A_\alpha + \id \big)^2
   = \big( \sum_{P_\omega \in \alpha} P_\omega \big)^2
   = \sum_{P_\omega \in \alpha} P_\omega^2
   + \sum_{P_\omega} \sum_{P_{\omega'}\neq P_\omega} P_\omega P_{\omega'}
\label{9} \end{equation}
\begin{displaymath}
  \Rightarrow \quad
  \textrm{Tr}A_\alpha^2 + 2 + N = (N+1) + (N+1)N\frac{1}{N}
  \quad \Leftrightarrow \quad
  \textrm{Tr}A_\alpha^2 = N \ .
\end{displaymath}
In our vector space a point face operator $A_\alpha$ corresponds to a
point on the ray from the origin through the center of the point
face. If continued in the other direction, this ray hits the center of
a facet, i.e. a face of dimension $N^2-2$. This facet is the convex
hull of the corners not included in the point face (like
\makebox{
$\triangle$ \hspace{-6.4mm}
\raisebox{-1.2ex}{$^\bullet\;^\bullet$}
$\triangle$ \hspace{-6.4mm}
\raisebox{-1.2ex}{$^\bullet\;^\bullet$}
$\triangle$ \hspace{-6.4mm}
\raisebox{-1.2ex}{$^\bullet\;^\bullet$}
$\triangle$ \hspace{-6.4mm}
\raisebox{-1.2ex}{$^\bullet\;^\bullet$}).}
Hence point faces and facets are in one-to-one correspondence, and
they are placed opposite to each other at the boundary of the
polytope. The case $N=2$ is special since its point faces are facets,
and conversely.

\begin{figure}
        \centerline{ \hbox{
                \epsfig{figure=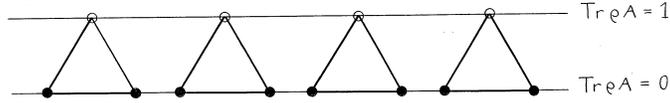,width=9cm}}} \vspace{3mm}
        \centerline{\parbox{10.4cm}{\caption{Sketch of the point
        face, corresponding facet, and planes of constant
        $\textrm{Tr}\rho A$, for $N=3$.} \label{MUB2}}}
\end{figure}

Every point face operator defines a set of parallel hyperplanes
through the equation
\begin{equation}
  \textrm{Tr}\rho A_\alpha = \textrm{constant} \ , 
\label{10} \end{equation}
where $\rho$ belongs to the set of Hermitian matrices of unit trace.
The facet lies in that hyperplane where the constant is zero, and the
point face where it is one (figure~\ref{MUB2}). We could have chosen
the point face operator $A_\alpha$ somewhere else on the ray, which
would just have given us other constants than zero and one at the
faces. The choice we made for $A_\alpha$ is in accordance with
Wotters' phase point operators~\cite{Wootters4}. All corners lies in
the hyperplanes with either $\textrm{Tr}\rho A_\alpha=0$ or
$\textrm{Tr}\rho A_\alpha=1$. Thus all points in the polytope have 
$0\leq\textrm{Tr}\rho A_\alpha\leq1$. Note that this observation
provides the formal proof that the faces really are faces, that is
that they consist of points belonging to the boundary of the
polytope. Note also that this very way of characterizing the facets 
of the polytope was used by Galv\~ao \cite{Galvao}, although he 
expressed it a little differently---his statement is that the discrete 
Wigner function, defined by Wootters \cite{Wootters4}, vanishes on 
facets of the polytope.

\begin{center}
 {\large \textbf{III. \ The affine planes}}
\end{center}

\noindent The discrete Wigner function used by
Wootters~\cite{Wootters1, Wootters4} is defined by
equation~(\ref{10})---the constant is the Wigner function
$\alpha$-component for the matrix $\rho$---for a selected subset of
$N^2$ face point operators.
Thinking about this construction from the point of
view of the complementarity polytope, we encounter a rather striking
geometrical fact. The idea is to ask whether it is possible to
inscribe a regular simplex---call it an \mbox{A-simplex}---in the
complementarity polytope, in such a way that its $N^2$ corners sit at
the centers of $N^2$ selected facets. We will turn this into a
combinatorial problem, that has a solution if and only if there exists
a finite affine plane of order $N$. Wootters' discrete Wigner function
lives on such an affine plane. But it is known that affine planes do
not exist for certain values of $N$, so we have found a geometrical
property of the polytope that definitely singles out certain values of
$N$ for special attention.  The argument relies on some elementary
facts taken from the branch of combinatorics that deals with finite
affine planes, Latin squares, and so on. A readable introduction that
contains all we need is the book by Bennett~\cite{Bennett}.

The question we will actually discuss is whether it is possible to
select $N^2$ of the matrices $A_{\alpha}$, such that they form an
\mbox{A-simplex}. Since these matrices lie on rays through the centers
of the facets, at a fixed distance from $\rho_*$, this question is
equivalent to the question as phrased above. These matrices will form
a regular simplex if and only if
\begin{equation} \mbox{Tr}A_{\alpha}A_{\beta} = N\delta_{\alpha\beta}
\ . \label{11} \end{equation}
This follows from the ratio $R_{N^2-1}^2/L^2=(N^2-1)/(2N^2)$ for
simplicies (see eq.~(\ref{vectors})). The radius of the
\mbox{A-simplex'} outsphere should be
$R_{N^2-1}=D(A_\alpha,\rho_\ast)$ and the side lengths
$L=D(A_\alpha,A_\beta)$. From this point on the index $\alpha$ is
restricted to run from $1$ to $N^2$. Observe that the matrices form an
orthogonal basis in the space of Hermitian matrices of order $N$,
i.e. in the Lie algebra of $U(N)$, albeit an unconventional one since
they all have unit trace.

To see how the requirement (\ref{11}) can be met, we modify equation
(\ref{9}) appropriately:
\begin{equation}
\mbox{Tr}A_{\alpha}A_{\beta}+2+N = 
\mbox{Tr}\sum_{P_\omega \in \alpha} P_{\omega} 
\sum_{P_\omega \in \beta} P_{\omega} \ . \end{equation}
Now we must choose the two sets of projectors $P_\omega$ that should
span the point faces $\alpha$ and $\beta$, such that the right hand side
yields $2+N$ (when $\alpha\neq\beta$). There are $N+1$ terms where the
two $P_\omega$s come from the same \mbox{P-simplex}. Such a term is
zero, according to equation~(\ref{P1}), unless we have the same choice
of $P_\omega$. Then it equals one ($\textrm{Tr}P_\omega^2=1$). The
remaining $N(N+1)$ terms, where the $P_\omega$s come from different
\mbox{P-simplices}, equal $1/N$, as in equation~(\ref{P2}), giving the
sum~$N+1$. Then we see that what we need is to choose exactly one
common projector $P_\omega$ in the sets spanning point faces $\alpha$
and $\beta$. And this must be done for every pair of the $N^2$
matrices that would give us an \mbox{A-simplex}. Hence we have a
combinatorial problem on our hands, and indeed a well known
combinatorial problem, as we will now proceed to show.

Order the point face operators in an $N\times N$ array. Let each point
face operator be represented by $N+1$ \mbox{P-simplices}, from which
exactly one out of $N$ corners is selected. In every column, the
choice of corner from the first simplex is the same, and in every row,
the choice of corner from the second simplex is the same. We
illustrate this first step of the construction for $N = 3$:
\begin{displaymath} \begin{array}{lllll} \triangle \hspace{-3.8mm} {\ }^\bullet 
\hspace{3.4mm} \triangle \hspace{-2.6mm} {\ }_{\bullet} \hspace{2mm} 
\triangle \ \triangle & \ 
& \triangle \hspace{-5mm} {\ }_{\bullet} \hspace{4mm} 
\triangle \hspace{-2.6mm} {\ }_{\bullet} \hspace{2mm} \triangle \ \triangle & 
\ & \triangle \hspace{-2.6mm} {\ }_{\bullet} \hspace{2mm} 
\triangle \hspace{-2.6mm} {\ }_{\bullet} \hspace{2mm} \triangle \ \triangle \\
\ & \ & \ & \ & \ \\
\triangle \hspace{-3.8mm} {\ }^\bullet 
\hspace{3.4mm} \triangle \hspace{-5mm} {\ }_{\bullet} \hspace{4.2mm} 
\triangle \ \triangle 
& \ & \triangle \hspace{-5mm} {\ }_{\bullet} 
\hspace{4.2mm} \triangle \hspace{-5mm} {\ }_{\bullet} \hspace{4.2mm} 
\triangle \ \triangle & \ &  
\triangle \hspace{-2.6mm} {\ }_{\bullet} \hspace{2mm} 
\triangle \hspace{-5mm} {\ }_{\bullet} \hspace{4.2mm} 
\triangle \ \triangle \\ 
\ & \ & \ & \ & \ \\
\triangle \hspace{-3.8mm} {\ }^\bullet 
\hspace{3.4mm} \triangle \hspace{-3.8mm} {\ }^\bullet 
\hspace{3.4mm} \triangle \ \triangle 
& \ & \triangle \hspace{-5mm} {\ }_{\bullet} \hspace{4mm} 
\triangle \hspace{-3.8mm} {\ }^\bullet 
\hspace{3.4mm} \triangle \ \triangle & \ & 
\triangle \hspace{-2.6mm} {\ }_{\bullet} \hspace{2mm} 
\triangle \hspace{-3.8mm} {\ }^\bullet 
\hspace{3.4mm} \triangle \ \triangle \end{array} 
\end{displaymath}

Now we must make choices of corners from the third simplex so that a
given choice occurs exactly once in every column and exactly once in
every row. If we think of the $N$ corners of a \mbox{P-simplex} as
letters in an alphabet of $N$ letters, we realize that our problem is
equivalent to that of finding a Latin square. Latin squares of order
$N$ always exist, and indeed there are many Latin squares. Altogether
we need $N-1$ Latin squares for the choices in the remaining
simplices. They have to obey consistency conditions that arise already
when we turn to the fourth simplex. Consider the pair of choices made
from the third and fourth simplex. There are $N^2$ such pairs
altogether, and they are not allowed to occur more than once in the
array---if they did, two different face point operators would have
more than one $P_{\omega}$ in common. This means that the third and
fourth simplex must define a pair of orthogonal Latin squares, where
two Latin squares are said to be orthogonal if every pair of letters,
one from each Latin square, defines a unique position in the
array. The same goes for every pair of the $N-1$ Latin squares that we
are trying to construct, so altogether we need $N-1$ mutually
orthogonal Latin squares. One can convince oneself that these choices
will give exactly one corner in common for every pair of point faces, 
as is required.

For $N = 3$, the finished construction looks as follows:
\begin{displaymath} \begin{array}{lllll} \triangle \hspace{-3.8mm} {\ }^\bullet 
\hspace{3.4mm} \triangle \hspace{-2.6mm} {\ }_{\bullet} \hspace{2mm} 
\triangle \hspace{-5mm} {\ }_{\bullet} \hspace{4mm} 
\triangle \hspace{-2.6mm} {\ }_{\bullet} \hspace{2mm} & \ 
& \triangle \hspace{-5mm} {\ }_{\bullet} \hspace{4mm} 
\triangle \hspace{-2.6mm} {\ }_{\bullet} \hspace{2mm} 
\triangle \hspace{-2.6mm} {\ }_{\bullet} \hspace{2mm} 
\triangle \hspace{-3.8mm} {\ }^\bullet \hspace{3.4mm} & 
\ & \triangle \hspace{-2.6mm} {\ }_{\bullet} \hspace{2mm} 
\triangle \hspace{-2.6mm} {\ }_{\bullet} \hspace{2mm} 
\triangle \hspace{-3.8mm} {\ }^\bullet \hspace{3.4mm} 
\triangle \hspace{-5mm} {\ }_{\bullet} \hspace{4.2mm} \\
\ & \ & \ & \ & \ \\
\triangle \hspace{-3.8mm} {\ }^\bullet 
\hspace{3.4mm} \triangle \hspace{-5mm} {\ }_{\bullet} \hspace{4.2mm} 
\triangle \hspace{-2.6mm} {\ }_{\bullet} \hspace{2mm} 
\triangle \hspace{-5mm} {\ }_{\bullet} \hspace{4.2mm} 
& \ & \triangle \hspace{-5mm} {\ }_{\bullet} 
\hspace{4.2mm} \triangle \hspace{-5mm} {\ }_{\bullet} \hspace{4.2mm} 
\triangle \hspace{-3.8mm} {\ }^\bullet 
\hspace{3mm} 
\triangle \hspace{-2.6mm} {\ }_{\bullet} \hspace{2mm} & \ &  
\triangle \hspace{-2.6mm} {\ }_{\bullet} \hspace{2mm} 
\triangle \hspace{-5mm} {\ }_{\bullet} \hspace{4.2mm} 
\triangle \hspace{-5mm} {\ }_{\bullet} \hspace{4.4mm} 
\triangle \hspace{-3.8mm} {\ }^\bullet \hspace{3.4mm} \\ 
\ & \ & \ & \ & \ \\
\triangle \hspace{-3.8mm} {\ }^\bullet 
\hspace{3.4mm} \triangle \hspace{-3.8mm} {\ }^\bullet 
\hspace{3.4mm} \triangle \hspace{-3.8mm} {\ }^\bullet \hspace{3.4mm} 
\triangle \hspace{-3.8mm} {\ }^\bullet \hspace{3.4mm} 
& \ & \triangle \hspace{-5mm} {\ }_{\bullet} \hspace{4mm} 
\triangle \hspace{-3.8mm} {\ }^\bullet \hspace{3.4mm} 
\triangle \hspace{-5mm} {\ }_{\bullet} \hspace{4mm} 
\triangle \hspace{-5mm} {\ }_{\bullet} \hspace{4mm} & \ &
\triangle \hspace{-2.6mm} {\ }_{\bullet} \hspace{2mm} 
\triangle \hspace{-3.8mm} {\ }^\bullet \hspace{3.4mm} 
\triangle \hspace{-2.6mm} {\ }_{\bullet} \hspace{2mm}  
\triangle \hspace{-2.6mm} {\ }_{\bullet} \hspace{2mm} 
\hspace{3.2mm} \end{array} 
\end{displaymath}

The conclusion is that our problem of inscribing a regular simplex (an
\mbox{A-simplex}) into the Complementarity Polytope is equivalent to
that of finding $N-1$ mutually orthogonal Latin squares. But the
latter problem is in its turn equivalent to that of defining a finite
affine plane of order $N$. These are combinatorial structures
containing $N^2$ points and $N+1$ pencils of $N$ parallel lines
each. (In the original quantum mechanics references, these are called
``foliations'' \cite{Wootters1} or ``striations'' \cite{Wootters4}.
However, ``pencils'' seems to be more standard \cite{Bennett}.) Each
of the lines contains $N$ points. Setting two of the pencils aside in
order to define an array of $N^2$ points, the axioms that define an
affine plane will require the final $N-1$ pencils to form mutually
orthogonal Latin squares. Hence we have proved that our problem is
equivalent to that of defining an affine plane of order $N$.

We have also set up a one-to-one correspondence between the
\mbox{P-simplices} that define the Complementarity Polytope, on the
one hand, and the pencils of parallel lines in an affine plane on the
other. This extends to a one-to-one correspondence between the corners
of our polytope and the lines of the affine plane. The points of the
latter correspond to a very special collection of $N^2$ facets of the
polytope. This should be obvious to readers familiar with finite
affine planes, but it is instructive to see how it goes for $N =
3$. In an affine plane, two points always determine a line. As an
example, the pair of points represented by 
$\triangle \hspace{-3.8mm} {\ }^\bullet
\hspace{2.4mm} \triangle \hspace{-3.8mm} {\ }^\bullet 
\hspace{2.4mm} \triangle \hspace{-3.8mm} {\ }^\bullet \hspace{2.4mm} 
\triangle \hspace{-3.8mm} {\ }^\bullet 
\hspace{3.4mm}$ 
and $\triangle \hspace{-5mm} {\ }_{\bullet} 
\hspace{3.8mm} \triangle \hspace{-5mm} {\ }_{\bullet} \hspace{3.8mm} 
\triangle \hspace{-3.8mm} {\ }^\bullet 
\hspace{2.6mm} \triangle \hspace{-2.6mm} {\ }_{\bullet} \hspace{2mm}$ 
(these are two point faces, belonging to the particular set of point 
faces that we have listed in our array) uniquely determine the line 
$\triangle \ \triangle \ \triangle \hspace{-3.8mm} {\ }^\bullet 
\hspace{2mm} \triangle $ (this is the only corner of the polytope 
that belongs to both faces). Also a pair of lines, such as 
$\triangle \ \triangle \ \triangle \hspace{-3.8mm} 
{\ }^\bullet \hspace{2mm} \triangle $ and $\triangle \ \triangle \ 
\triangle \ \triangle \hspace{-3.8mm} {\ }^\bullet \hspace{3.4mm}$, 
uniquely determine a point, in this case 
$\triangle \hspace{-3.8mm} {\ }^\bullet
\hspace{2.4mm} \triangle \hspace{-3.8mm} {\ }^\bullet 
\hspace{2.4mm} \triangle \hspace{-3.8mm} {\ }^\bullet \hspace{2.4mm} 
\triangle \hspace{-3.8mm} {\ }^\bullet 
\hspace{3.0mm}$
(since this is the only point face in our array that contains this
particular pair of corners). In the latter case we must make an
exception if we pick two corners from the same P-simplex; an example
would be
$\triangle \ \triangle \ \triangle \hspace{-3.8mm} {\ }^\bullet 
\hspace{2mm} \triangle $ and $\triangle \ \triangle \
\triangle \hspace{-2.6mm} {\ }_{\bullet} \hspace{1.2mm} \triangle $. 
They represent lines coming from the same pencil of parallel lines.

About the existence of affine planes, the following is known: Affine
planes of order $N$ do exist if $N = p^k$, where $p$ is a prime
number. They do not exist if $N = 4k + 1$ or $N = 4k + 2$, unless
possibly when $N$ is the sum of two squares (a statement known as the
Bruck-Ryser theorem). They do not exist if $N = 10$ (this case has
been settled by state-of-the-art modern computers).  All other cases,
beginning with $N = 12$, are open (so much for today's
computers). Concerning the cases where affine planes exist, we observe
that they can be constructed using ordinary vector space methods, if
the scalar numbers employed are taken from finite number fields. This
option is open only if $N = p^k$, but examples of affine planes that
do not arise in this way are known---thus, for $N = 9$, four
nonisomorphic affine planes are known, only one of which can be
coordinatized by a field. However, all examples so far constructed have 
$N=p^k$.

It will not have escaped the reader that we have not explained to what 
extent the question of the existence of an affine plane has anything to 
do with the question of the existence of a complete set of MUBs. We will make 
some comments on this issue in the concluding section---but no 
explanation will be forthcoming.

\begin{center}
 {\large \textbf{IV. \ Is there some strangeness in the proportions?}}
\end{center}

\noindent We have two convex bodies in dimension $N^2-1$: the set of density
matrices and the complementarity polytope. Now we study some more
geometrical facts about these sets.

Let us derive the volume of the complementarity polytope. In every
$N-1$ dimensional subspace containing a \mbox{P-simplex}, the
coordinates of the $N$ corners are given by the vectors
\begin{equation} \label{vectors}
  \left[ \begin{array}{c}
            -r_1 \\ -r_2 \\ -r_3 \\ \vdots \\ \!\! -r_{N-1} \!
         \end{array} \right]\ \
  \left[ \begin{array}{c}
            R_1 \\ -r_2 \\ -r_3 \\ \vdots \\ \!\! -r_{N-1} \!
         \end{array} \right] \ \
  \left[ \begin{array}{c}
            0 \\ R_2 \\ -r_3 \\ \vdots \\ \!\! -r_{N-1} \!
         \end{array} \right] \ \  \cdots \ \
  \left[ \begin{array}{c}
            0 \\ \vdots \\ 0 \\ R_{N-2} \\ \!\! -r_{N-1} \!
         \end{array} \right] \ \\
  \left[ \begin{array}{c}
            0 \\ \vdots \\ \vdots \\ 0 \\ R_{N-1}
         \end{array} \right] \ \
\end {equation}
where $r_n = 1/\sqrt{2n(n+1)}$ and $R_n = n r_n$ are the radii of the
inscribed and the circumscribed spheres of an $n$ dimensional
simplex. From this we get the $N(N+1)$ vectors for the corners of the
polytope, in the full $N^2-1$ dimensional space. The polytope consists
of cones with the facets as their bases. If we choose $N-1$ vectors
from each \mbox{P-simplex}---this can be done by discarding the first
of the vectors above (in eq.~(\ref{vectors}))---we get $N^2-1$ vectors
such that they span a cone with a facet as its base. The volume of the
cone is found from the determinant of the spanning vectors, and the
volume of the polytope by multiplying with the number of facets,
$N^{N+1}$:
\begin{equation}
  V_{polytope} = N^{N+1} \frac{1}{(N^2-1)!} (R_1 R_2 \cdots R_{N-1})^{N+1} \\
  = \frac{ \sqrt{N}^{N+1} }{ (N^2-1)! \sqrt2^{N^2-1} } \ .
\end{equation}

How does this volume compare to the volume of the set of density
matrices? This second volume is given by~\cite{Karol}
\begin{equation}
  V_{\{\rho\}} = \frac{ \sqrt{N} \pi^{\frac{N(N-1)}{2}} }
                      { \sqrt2^{N-1} }
                 \frac{ 1!\,2!\,3! \cdots (N-1)! }
                      { (N^2-1)! } \ .
\end{equation}
(A factor $1/\sqrt2$ for every dimension is included, to adjust for
our different conventions in the definition of distance.)
Hence, the ratio between the volumes of the polytope and the set of
density matrices is
\begin{equation}
  \frac{ V_{polytope} }{ V_{\{\rho\}} }
  = \frac{ \sqrt{N}^N }{ \sqrt{2\pi}^{N(N-1)} }
    \frac{ 1 }{ 1!\,2!\,3! \cdots (N-1)! } \ .
\end{equation}
For $N=2$ we get $V_{polytope}=1/6$ and $V_{\{\rho\}}=\pi/6$, as we
should for an octahedron and the Bloch ball, and
$V_{polytope}/V_{\{\rho\}}=1/\pi$. The fraction
$V_{polytope}/V_{\{\rho\}}$ is less than one and decreases rapidly
with the dimension~$N$---a reasonable result remembering that most of
the volume in a higher dimensional ball lies near the surface: The set
of density matrices touches the circumscribed sphere at all the pure
states, but the polytope only at its corners. Also, the
inscribed sphere of the set of density matrices has a larger radius
than the inscribed sphere of the polytope. So the \emph{volume} of the
polytope is certainly not too large to be included in the set of
density matrices. But this doesn't say much about whether the
\emph{polytope} fits inside the set of density matrices or not.

We also compute the radius of the insphere and the area of the
polytope. We can get a point on the insphere of the polytope as an
equal weight convex combination of points on the inspheres of the
\mbox{P-simplices}, one from each simplex. The radius of the insphere
of a simplex in $N-1$ dimensions is $r_{N-1}=1/\sqrt{2N(N-1)}$. The
radius $r_{in}$ of the insphere of the whole polytope, built of $N+1$
simplices, is then (by Pythagoras' theorem and remembering that the
\mbox{P-simplices} are totally orthogonal)
\begin{equation}
  r_{in} = \sqrt{(N+1) \left( \frac{r_{N-1}}{N+1}\right)^2}
         = \frac{1}{\sqrt{2N(N^2-1)}} \ .
\end{equation}

The area of the polytope is the ($N^2-2$ dimensional) area of its
facets. This is, again, the bases of the cones building up the
polytope. The height of the cones is the radius $r_{in}$. Using
$V_{cone}=A_{base}r_{in}/\textrm{dim}(cone)$ we get
\begin{equation} \label{A/V}
  \frac{ A_{polytope} }{ V_{polytope} }
  = \frac{\textrm{dim}(polytope)}{r_{in}} = \sqrt{2N(N^2-1)}(N^2-1)
\end{equation}
and
\begin{equation}
  A_{polytope}
  = \frac{ \sqrt{N}^{N+2} \sqrt{N^2-1} }{ (N^2-1)! \sqrt{2}^{N^2-2} } \ .
\end{equation}
Multiplying equation (\ref{A/V}) with the radius of the outsphere,
$R=R_{N-1}=\sqrt{N-1}/\sqrt{2N}$, to fix the scale, we then compare
with the set of density matrices:
\begin{equation}
  R\frac{ A_{polytope} }{ V_{polytope} }
  = (N-1)^{\frac{1}{2}}(N^2-1)^{\frac{3}{2}}
  \quad , \quad
  R\frac{ A_{\{\rho\}} }{ V_{\{\rho\}} }
  = (N-1)(N^2-1) \ . \label{20}
\end{equation}
Asymptotically, this ratio behaves like $(\mbox{dimension})^{7/4}$ for the 
polytope, and like $(\mbox{dimension})^{3/2}$ for the set of density matrices. 
The latter value of the exponent happens to coincide with that for hypercubes, 
while regular simplices behave like $(\mbox{dimension})^2$. So the polytope is 
closer to the simplex---but we can see nothing that singles out some particular 
values of $N$. 

One thing that characterizes convex bodies is their largest inscribed
ellipsoid. For example whether this ellipsoid turns out to be a ball
(corresponding to the insphere) or not. If it is not a ball the body
is kind of stretched or flattened in some directions. Here we can rely
on a useful theorem due to Fritz John~\cite{John}.

\emph{John's theorem:} \\
\emph{The ball $\mathcal{B}$ is the ellipsoid of maximal
volume that can be contained in a convex body $K$ if and only if:
$\mathcal{B} \in K$ and there exist vectors $\{ {\bf u}_i \}_1^m$ at
the boundary of both $\mathcal{B}$ and $K$, and a set of positive
numbers $\{c_i\}_1^m$, such that\\
\begin{equation} \label{John}
  (i)  \qquad \sum_{i=1}^m c_i{\bf u}_i = 0 \qquad \textrm{and} \qquad
  (ii) \qquad \sum_{i=1}^m c_i{\bf u}_i \otimes {\bf u}_i = \id \ .
\end{equation}}

These conditions correspond to the possibility to have masses $c_i$ at
some points ${\bf u}_i$ where the sphere touches the boundary of the
convex body, such that the center of mass is at the center of the
sphere, and such that the inertia tensor is the identity.
 
Are the inspheres of the set of density matrices and of the
complementarity polytope the largest inscribed ellipsoids?

\begin{figure}
        \centerline{ \hbox{
                \epsfig{figure=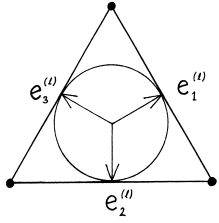,width=4cm}}}  \vspace{-3mm}
        \centerline{\parbox{10.4cm}{\caption{The vectors ${\bf
        e}_k^{(l)}$ in one of the \mbox{P-simplices}, for $N=3$.}
        \label{MUB3}}}
\end{figure}

Let's start with the density matrices. In those dimensions where we
can find a complete set of $N+1$ MUBs, look at the corresponding
\mbox{P-simplices}. The insphere of the set of density matrices has
the same radius as the inspheres of the simplices
\cite{Harriman}. Anti-parallel to the vectors for the $N$ corners of
each simplex there are vectors corresponding to the points where the
insphere touches the boundary---call them ${\bf e}_k^{(l)}$, where
$l=1,\dots,N+1$ labels the simplices and $k=1,\dots,N$ labels the
different vectors in each simplex (figure~\ref{MUB3}). Then
\begin{equation}
  \sum_{k=1}^N {\bf e}_k^{(l)} = 0 \ \ \textrm{ hence } \ \ 
  \sum_{l=1}^{N+1} \sum_{k=1}^N {\bf e}_k^{(l)} = 0 \ ,
\end{equation}
and the first condition in the theorem is fulfilled (for $c_i=c$, any
constant). In the second condition, for every simplex, the sum over
$k$ gives the identity (times a factor) in the $N-1$ dimensional
subspace of the simplex. And then the sum over $l$ gives the identity
in the full $(N+1)(N-1)$ dimensional space:
\begin{equation} \label{tensor}
  \sum_{k=1}^N {\bf e}_k^{(l)} \otimes {\bf e}_k^{(l)} \propto \id_{N-1}
   \ \ \textrm{ hence } \ \ 
  \sum_{l=1}^{N+1} \sum_{k=1}^N {\bf e}_k^{(l)} \otimes 
                                {\bf e}_k^{(l)} \propto \id_{N^2-1} \ .
\end{equation}
Thereby we have proved that the insphere of the set of density
matrices is the largest inscribed ellipsoid, when we have a complete
set of MUBs.

What can we say about the largest ellipsoid in the set of density
matrices, without using a complete set of MUBs? We are looking for
points that fulfill John's theorem. These points should lie at the
insphere and at the boundary. The possible points are those that sit
opposite to the pure states. (The innermost states at the boundary
have spectrum $(0,1/N,\dots,1/N)$ and lie opposite to the states with
spectrum $(1,0\dots,0)$, i.e. the pure states, lying at the
circumscribed sphere.) Thus we can look at the pure states
$|\Psi\rangle$ instead (in the theorem it only corresponds to a change
of the constants $c_i$, by a factor).

A vector ${\bf u}$ corresponds to a density matrix
$|\Psi\rangle\langle\Psi|$. The first condition (in eq.~(\ref{John}))
becomes
\begin{equation}
  \sum_i c_i |\Psi_i\rangle\langle\Psi_i| \propto \id_N \ .
\end{equation}
This is the condition that the set of projectors
$\{|\Psi_i\rangle\langle\Psi_i|\}$ form a POVM (with one dimensional
projectors as its element). A first requirement from the second
condition,
\begin{equation}
  \sum_i c_i{\bf u}_i \otimes {\bf u}_i = \id_{N^2-1} \ ,
\end{equation}
is that the states must span the whole space of density
matrices. Since ${\bf u}_i \otimes {\bf u}_i$ is a projector of rank
one, we must have at least $N^2-1$ vectors ${\bf u}_i$. But if we have
only $N^2-1$ vectors, they have to be linearly independent and the
first condition can not be fulfilled. Therefore we must have at least
$N^2$ vectors. One set of vectors that would suit us---if they can be
found among the pure states---is a set of vectors spanning an $N^2-1$
dimensional regular simplex (then all $c_i=c$, a constant). This is exactly
what have been called a \emph{symmetric informationally complete POVM}
(SIC-POVM). These have been found in dimensions
$N=1,3,4,5$~\cite{Zauner,SIC-POVM} and 6~\cite{Grassl}, and it has been
conjectured that they exist in all dimensions (based on numerical
solutions up to $N=45$)\cite{SIC-POVM}.  Thus, if there exists a
SIC-POVM, then there is no larger inscribed ellipsoid in the set of
density matrices than the insphere. It is interesting to note that a
SIC-POVM is found also for the case $N=6=2\cdot3$---an $N$ which is
not a power of a prime.

Just one more comment on John's theorem and density matrices. There
could be many other ways to fulfill the conditions, than using a
SIC-POVM, since we can use more vectors ${\bf u}_i$ and we have the
freedom to use different weights $c_i$. Altogether we find it probable
that the largest ellipsoid in the set of density matrices is always a ball.

Now we turn to the complementarity polytope. The points at which the
insphere of the polytope touches the boundary are convex combinations
of vectors ${\bf e}_k^{(l)}$, one from each of the $N+1$
\mbox{P-simplices}:
\begin{equation}
  {\bf u}_{k_1,\dots,k_N} = \frac{1}{N+1} \sum_{l=1}^{N+1} {\bf e}_{k_l}^{(l)} \ ,
\end{equation}
for every combination of values of $k_1,\dots,k_N$.

Since the sum over the vectors ${\bf e}_k^{(l)}$ for each simplex
(given~$l$) is zero, the sum over all vectors ${\bf
u}_{k_1,\dots,k_N}$ will also be zero, thus the first condition in the
theorem is fulfilled (for $c_i=c$ any constant).

What about the second condition?
\begin{equation}
  \sum_{k_1,\dots,k_N} {\bf u}_{k_1,\dots,k_N} \otimes {\bf u}_{k_1,\dots,k_N} 
  \propto
  \sum_{k_1,\dots,k_N} \sum_{l=1}^{N+1} {\bf e}_{k_l}^{(l)}
               \otimes \sum_{l=1'}^{N+1} {\bf e}_{k_l'}^{(l')}
\end{equation}
First look at the terms where $l\neq l'$, and for a given $l=L$.
\begin{equation}
  \sum_{k_1,\dots,k_N} {\bf e}_{k_L}^{(L)}
               \otimes \sum_{  \begin{array}{c}
                               {\scriptstyle l=l'} \\[-5pt]
                               {\scriptstyle l\neq L}
                               \end{array}}
                       {\bf e}_{k_l'}^{(l')} =
  \Big( \sum_{k_L} {\bf e}_{k_L}^{(L)} \Big) \otimes
  \Big( \hspace{-8pt} \sum_{   \begin{array}{c}
                               {\scriptstyle k_1,\dots,k_{L-1}} \\[-5pt]
                               {\scriptstyle k_{L+1},\dots,k_N}
                               \end{array}}
        \hspace{-8pt} \sum_{  \begin{array}{c}
                               {\scriptstyle l=l'} \\[-5pt]
                               {\scriptstyle l\neq L}
                               \end{array}}
         {\bf e}_{k_l'}^{(l')} \Big) = 0
\end{equation}
This is zero since the first factor, the sum over the vectors for one
\mbox{P-simplex}, is zero. Then all that is left are the terms
where~$l=l'$:
\begin{equation}
  \sum_{k_1,\dots,k_N} \sum_{l=1}^{N+1} {\bf e}_{k_l}^{(l)}
               \otimes {\bf e}_{k_l}^{(l)} =
  \sum_{l=1}^{N+1} \sum_{k=1}^N {\bf e}_k^{(l)} \otimes 
                                {\bf e}_k^{(l)} \propto \id_{N^2-1} \ ,
\end{equation}
as in equation~(\ref{tensor}). Thus the second condition of the
theorem is fulfilled, and the insphere is the largest ellipsoid that
can be inscribed in the polytope. In this sense, neither the set of density
matrices nor the complementarity polytope is ``skew''.

\begin{center}
 {\large \textbf{V. \ Conclusions}}
\end{center}

\noindent The Complementarity Polytope exists in any vector space of
dimension ${N^2-1}$. The question whether a complete set of $N+1$ MUBs
exists has been much discussed recently; in addition to the references
that we have already given, see
refs.~\cite{Zeilinger,Vatan,Rubin,Durt}. It is equivalent to the
question whether the Complementarity Polytope can be inscribed in the
convex body of density matrices. If this is indeed possible only for
some special values of $N$, say $N = p^k$ where $p$ is prime, then it
seems to us that one can reasonably expect to find some strangeness in
the proportions of, either the body of density matrices, or the
Complementarity Polytope, for these values of $N$. We have examined
the latter in section IV, with regard to its volume, area and the
shape of the maximal inscribed ellipsoid, but we were unable to put
our finger on any peculiarities. The prime factorization of $N$ does
not appear in these calculations at all. In section III we did find a
peculiarity, namely that a regular simplex can be inscribed in the
polytope (with its corners sitting at the midpoints of $N^2$ selected
facets) if and only if an affine plane of order $N$ exists. This
condition excludes an infinite number of non-prime values of $N$, but
unfortunately we are unable to see what, exactly, this has to do
with the existence of MUBs. What we do know is that an affine plane 
coordinatized by a field can be used to derive the existence of 
MUBS \cite{Wootters4}, but we do not know anything about the converse. 

It remains possible that there is a connection between the existence
of affine planes and the existence of MUBs~\cite{SPR}. In fact, as
noted by Zauner~\cite{Zauner}, it looks suspicious that only 3 MUBs
have been found for $N = 6$, where it is known to be impossible to
find a single pair of orthogonal Latin squares. Therefore we believe
that our underlying idea is sound, that the translation of the affine
plane question into a geometric question about the Complementarity
Polytope is interesting, and that further investigation of the latter
may bear fruit.

\begin{center}
 {\large \textbf{Acknowledgements}}
\end{center}

\noindent We thank Gunnar Bj\"ork, Bengt Nagel and Ninos Poli for 
discussions (over the years), and Karol \.Zyczkowski for the remark 
following equation~(\ref{20}).

\end{document}